\documentclass[fleqn,usenatbib]{mnras}

\usepackage{newtxtext,newtxmath}

\usepackage[T1]{fontenc}
\usepackage{ae,aecompl}

\usepackage{amsmath}	
\usepackage{amssymb}	

\usepackage{ulem}  
\usepackage{color} 

\usepackage[flushleft]{threeparttable}

\usepackage{graphicx} 
\graphicspath{{figs/}}
\usepackage{rotating} 
\usepackage{caption}
\usepackage{capt-of}
\usepackage{tabularx,colortbl}

\def\aap{A\& A}

\def\aj{{AJ}}
\def\apj{ApJ}
\def\apjl{ApJL}
\def\apjs{ApJS}

\def\araa{{ARA\&A}}

\def\mnras{MNRAS}

\def\pasp{{PASP}}

\def\alamenos#1{$^{-#1}$}

\def\diezala#1{10$^{#1}$}
\def\kms{{km~s\alamenos 1}}
\def\ppcc{{\rm cm}^{-3}}

\title[Dispersion of clusters via gravitational feedback] {Flipping-up the field: gravitational feedback as a mechanism for young clusters dispersal}

 \author[Zamora-Avil\'es et al. ]{Manuel Zamora-Avil\'es,$^{1,2}$\thanks{E-mail: mzamora@inaoep.mx}  
   Javier Ballesteros-Paredes,$^2$  Jes\'us Hern\'andez,$^3$ 
   \newauthor Carlos Rom\'an-Z\'u\~{n}iga,$^3$ Ver\'onica Lora,$^2$ and Marina Kounkel$^4$ \\ 
   $^1$CONACYT-Instituto Nacional de Astrof{\'i}sica, {\'O}ptica y 
    Electr{\'o}nica, Luis E. Erro 1, 72840 Tonantzintla, Puebla, M{\'e}xico \\
   $^2$Instituto de Radioastronom{\'i}a y Astrof{\'i}sica, Universidad Nacional Aut{\'o}noma de M{\'e}xico, Apdo. Postal 72-3 (Xangari), \\
   Morelia, Michoc{\'a}n 58089, M{\'e}xico \\
   $^3$Instituto de Astronom\'ia, Universidad Nacional Aut{\'o}noma de 
     M{\'e}xico, Unidad Acad{\'e}mica en Ensenada, Ensenada 22860 BC, M{\'e}xico \\
   $^3$Department of Physics and Astronomy, Western Washington University, 516 High St., Bellingham, WA 98225, USA}

\date{Accepted XXX. Received YYY; in original form ZZZ}

\pubyear{2019}

\begin{document}
\label{firstpage}
\pagerange{\pageref{firstpage}--\pageref{lastpage}}
\maketitle

\begin{abstract}

Recent analyses of Gaia data have provided direct evidence that most young stellar clusters are in a state of expansion, with velocities of the order of $\sim$~0.5~\kms. Traditionally, expanding young clusters have been pictured as entities that became unbound due to the lack of gravitational binding once the gas from the parental cloud that formed the cluster has been expelled by the stellar radiation of the massive stars in the cluster.  In the present contribution we used radiation-magnetohydrodynamic numerical simulations of molecular cloud formation and evolution to understand how stellar clusters form and disperse. We found that the ionising feedback from the newborn massive stars expels the gas from the collapse centre, flipping-up the gravitational potential as a consequence of the mass removal from the inside-out. Since neither the parental clouds, nor the formed shells are distributed symmetrically around the H\,{\normalsize II} region, net forces pulling out the stars are present, accelerating them towards the edges of the cavity. We call this mechanism ``gravitational feedback", in which the gravity from the expelled gas appears to be the crucial mechanism producing unbound clusters that expand away from their formation centre in an accelerated way in young stellar clusters. This mechanism naturally explains the "Hubble flow-like" expansion observed in several young clusters.

\end{abstract}



\section{Introduction} \label{sec:intro} 

\begin{figure}
\includegraphics[width=1.0\hsize]{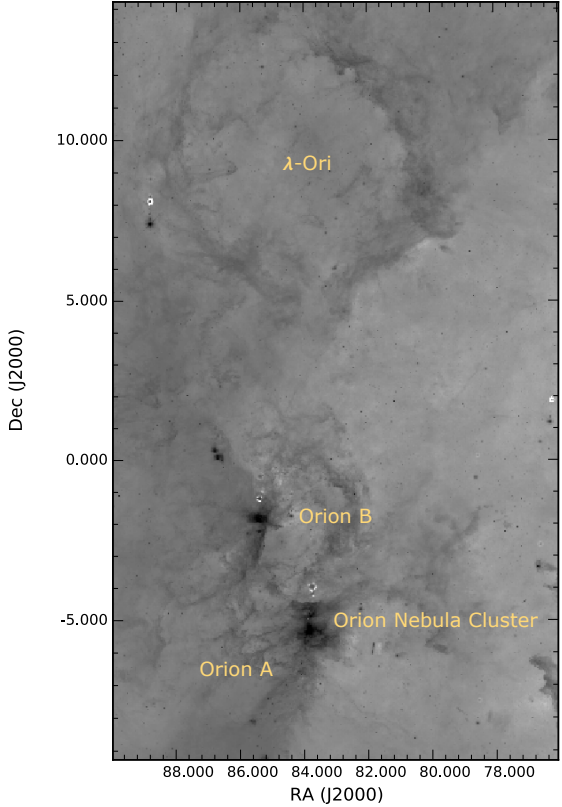} 
\caption{A 12 $\mu$m image of the Orion Complex from the Wide-field Infrared Survey Explorer (WISE) Full-sky High-resolution Atlas of Dust Emission \citep[WSSA;][]{wssa13}. The map reveals the structure of bubbles in the Orion Complex, which is particularly clear in $\lambda$-Ori. As a reference, an angular size of 1$^\circ$ corresponds to $\sim$7 pc (assuming a distance of 400 pc to the Orion MC complex).
\label{fig:orionextinction}}
\end{figure}

In almost any direction with significant H\,{\normalsize I} 21~cm emission, the interstellar medium exhibits cavities in the gas. They can be easily detected as shell or ring-like structures in  emission-line maps tracing either neutral atomic  \citep[e.g.,][see also Fig.~\ref{fig:orionextinction}]{Wade57, HartmannBurton97}, ionised \citep[optical H$_\alpha$,  e.g,][]{Madsen06} or molecular hydrogen \citep[through CO emission or infrared dust emission, for instance; e.g.,][]{Maddalena+87, Lang+00}. These systems have been systematically identified as expanding structures produced by a combination of radiative and mechanical feedback effects from massive stars, such as ionising radiation, winds, supernovae explosions, or collisions with high-velocity clouds \citep{Tenorio-Tagle+88}.

Recently, \citet{Kounkel+18} and \cite{Kuhn+19} analysed proper motions from Gaia-DR2 of young stellar clusters, and concluded that at least 75\% are expanding systems, with the most statistically significant expanding cases were $\lambda-$Ori in the first work, and NGC 6530 and Cep B in the second. These cases exhibit ``Hubble flow-like" expansion signatures, i.e., larger velocities far from the centre of the bubble. 
\citet{RZ2019a} showed that Hu\-bble\--\-li\-ke 
ex\-pan\-sion of young stellar groups can be observed at the scales of star forming complexes ($10^2$-$10^3$ pc).

\begin{figure*}
\includegraphics[width=0.49\hsize]{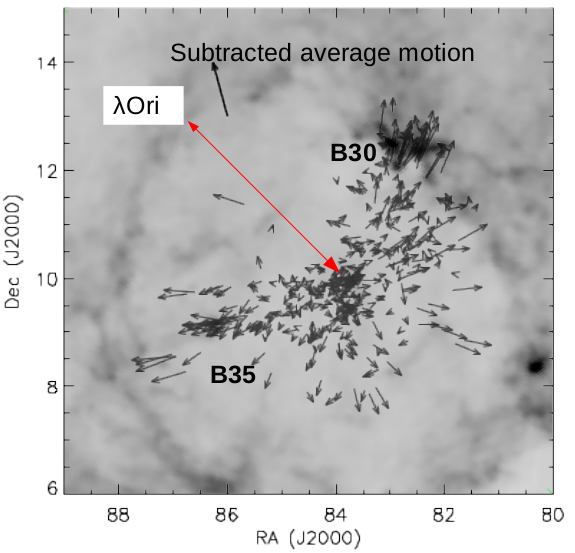}
\includegraphics[width=0.49\hsize]{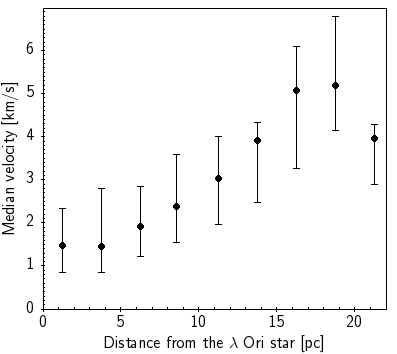}
\caption{{{\it Left:} Dust emission map of the $\lambda$-Ori region \citep{schlegel1998} in which we have overplotted the velocity vectors of the stars deduced by proper motions from the GAIA-DR2 survey. {\it Right:} Median value of the proper motion modulus as a function of the distance from the $\lambda$-Ori star. We estimate these velocities from the median proper motion modulus assuming a distance of 400 pc \citep{kounkel17}. Error bars represent the first and third quartiles of the bin. \label{fig:propermotions}}}
\end{figure*}

The particular case of $\lambda$-Ori (see Fig.~\ref{fig:orionextinction}) is quite inte\-res\-ting. This region is located at the northwest end of the Orion Complex at a distance of 400$\pm$20 pc (Mathieu, 2008; Kounkel et al., 2018) 
It shows a very symmetric ring-like structure\footnote{Although the morphology of $\lambda$-Ori could be more complicated \citep{Lee+15}, the symmetric ring is ``by eye" the more prominent structure.} of gas and dust with a radius of $\sim$20~pc that includes a central cluster (Collinder 69) with an age between $\sim${5 Myr (e.g, Murdin \& Penston 1977; Lang et al. 2000)}  and 10 Myr \citep{Bell+13} and several younger stellar regions (B30, B35, LDN1588 and LDN1603) near the external ring. For the central cluster, the projected velocity vectors tend clearly to point away from the centre of the ring  (see Fig. \ref{fig:propermotions}). 

Two scenarios have been proposed for the formation and expansion of this region. One of them, presented by \citet{Maddalena+87}, associates the expansion to transfer of momentum from the H\,{\normalsize II} region of $\lambda$-Ori. The other one, by \citet{CunhaSmith96}, assumes that the $\lambda$-Ori ring is a supernova remnant that formed the structure and disrupted the cluster. Although this scenario is frequently quoted in the literature as the mechanism that could have produced this region, there is no substantial evidence of the SN progenitor. Some candidates have been proposed and discarded, e.g. the Geminga pulsar, \citep{Pellizza+05}, or the $\gamma$-ray pulsar J0357+3205, \citep{Kirichenko+14}. The SN conjecture and its consequences for the past and future evolution of $\lambda$-Ori  has been explored by \citet{DolanMathieu01, DolanMathieu02} who propose a progenitor of 30-40 $M_\odot$ and the explosion event at about 1 Myr ago, coinciding with the halt of star formation. 

By combining parallaxes and proper motions from Gaia-DR2 \citep{gaiadr2a,gaiadr2b} with radial velocities from the APOGEE-YSO survey \citep{zasowski17}, kinematic members of the clusters B30, B35 and  Collinder 69, were identified, revealing the projected motion of stars in $\lambda$-Ori with exquisite detail \citep{Kounkel+18}. In the left panel of  Fig.~\ref{fig:propermotions} we overplot a map of dust infrared emission in this region \citep{schlegel1998} along with proper motions of kinematic members. The right panel of the same figure shows the median value of the velocity\footnote{The velocity values reported in this plot are only lower limits, since it has not been taken into account the perspective effects, i.e., the contribution of the positive radial velocity of the cluster in lowering of the proper motions.} as a function of the distance to a centre defined by the position of the $\lambda$-Ori star. As can be seen, proper motions not only tend to point outwards, but the magnitude of the projected velocities is clearly larger for the stars that are located further away from the centre of the cluster. This behaviour is consistent with the ``free expanding" clusters reported by \cite{Kuhn+19}.

A possible scenario for understanding the kinematical characteristics of the stellar cluster shown in Fig.~\ref{fig:propermotions} could be one in which the cluster is formed with a given velocity dispersion, and becomes unbound once the gas is evacuated from the centre. In this case, the stars with larger velocities reach larger distances. For $\lambda-$Ori, if its stars have been flying apart from the centre at a constant velocity, the ages of the stars in the periphery should be close to $\tau\sim$~20~pc/6~\kms$ \sim$3~Myr.\footnote{We assume that the velocity vectors estimated in Fig.~\ref{fig:propermotions} for the periphery are underestimated by a $\sim$1~km/sec, due to the perspective effects of a receding  cluster, provided the distance to the region (400 pc) and the recession velocity ($\sim$20 \kms).} This number is smaller than the estimated ages of about $\sim 5-$ 10~Myr for the central $\lambda$-Ori cluster \citep[e.g.,][]{Murdin+77, Bell+13},  and thus, it cannot be considered that these stars were formed in the  central cluster. On the other hand, it is more likely that the stars in the periphery were formed in the already expanding shell. Indeed, the ages of the B30 and B35 regions are about $\sim$2-3~Myr \citep{mathieu2008,barrado+2018}, and thus, these stars could be formed when the shell was half of the size, and thus the stars will inherit the velocity of their parent clump.

An alternative mechanism for the velocity structure presenting ``Hubble-like" flows, which has being unforeseen in the literature is the possibility that the removal of mass due to the expansion of the H\,{\normalsize II} region not only removes the gravitational potential, as has being studied previously \citep[e.g.,][]{Smith+13, Farias+18}, but it actually flips it up, generating a cusp at the position of the H\,{\normalsize II} region, and moving away the potential well to the outside. This reversal of the gravitational field shape will add an outward acceleration to the stars in the clusters.

Motivated, thus, by the morphological and kinematical features of $\lambda-$Ori, and in order to understand the dynamics of young stellar clusters, in this contribution we analyse a numerical simulation, which follows the evolution of a magnetised molecular cloud from its creation from a compressed warm neutral medium, to its destruction by ionising feedback from massive stars. We find that the gravitational potential of the expelled gas (powered by ionisation feedback) is crucial to unbound the central clusters. In \S\ref{sec:methods} we describe the numerical method. In \S\ref{sec:results} we present our results, describing the gravitational acceleration experienced by the stellar particles in the simulations. In \S\ref{sec:discussion} we discuss our results, and in \S\ref{sec:summary} we summarise our main conclusions.

\section{Numerical methods}\label{sec:methods}

We analysed a numerical radiation-magnetohydrodynamic  simulation presented by Zamora-Avil\'es et al. (2009; see also Zamora-Avil\'es et al. 2018),
carried out with the the Eulerian adaptive mesh refinement code {\tt FLASH} \citep[v2.5;][]{flash} which was aimed at studying the structure and expansion laws of H\,{\normalsize II} regions evolving in highly structured and collapsing MCs. This simulation includes the more relevant physical processes for cloud formation and evolution, such as self-gravity, magnetic fields, heating and cooling, sink formation, and ionising feedback.

The MC is formed by two warm neutral streams colliding at the centre of the numerical box. The compressed layer is cold due to a phase transition triggered by the thermal instability. At the same time, the inflows inject the characteristic turbulence observed in MCs through some dynamical instabilities \citep[see, e.g.,][]{Heitsch+06}. As the cold layer continues accreting material from the inflows, it increases its mass and column density, allowing the formation of molecular gas, and eventually the entire cloud enters in a state of {\it global hierarchical collapse}, in which the smaller scales collapse first due to their shortest free-fall times \citep[e.g.,][]{VS+19}. Once massive stars appear in our highly structured clouds, their ionising radiation generate over-pressured H\,{\normalsize II} regions that push away the dense gas, thus disrupting the more massive collapse centres and regulating the star formation activity.

The warm, neutral streams are cylindrical, each with 32 pc in radius and 112 pc in length along the $x$-direction, containing warm gas in thermal equilibrium with density $n_0=2 \, {\rm cm}^{-3}$ and temperature $T_0=1450$~K. These streams are completely contained in a numerical box of size 256 pc in the $x$-axis and 128 pc in the $y$-, $z$-axes. The streams collide at a transonic velocity of $\sim 7.5 \, {\rm km~s}^{-1}$. We impose a background turbulent velocity field (with Mach number of 0.7, subsonic in the warm gas) in order to trigger the dynamical instabilities in the shocked layer. The magnetic field is initially uniform along the $x$-direction with a strength of $3 \, \mu$G. This corresponds to an initial mass-to-flux ratio of 1.5 times the critical value, and thus the cloud is magnetically super-critical as a whole. 

\begin{figure*}
\includegraphics[width=1.0\hsize]{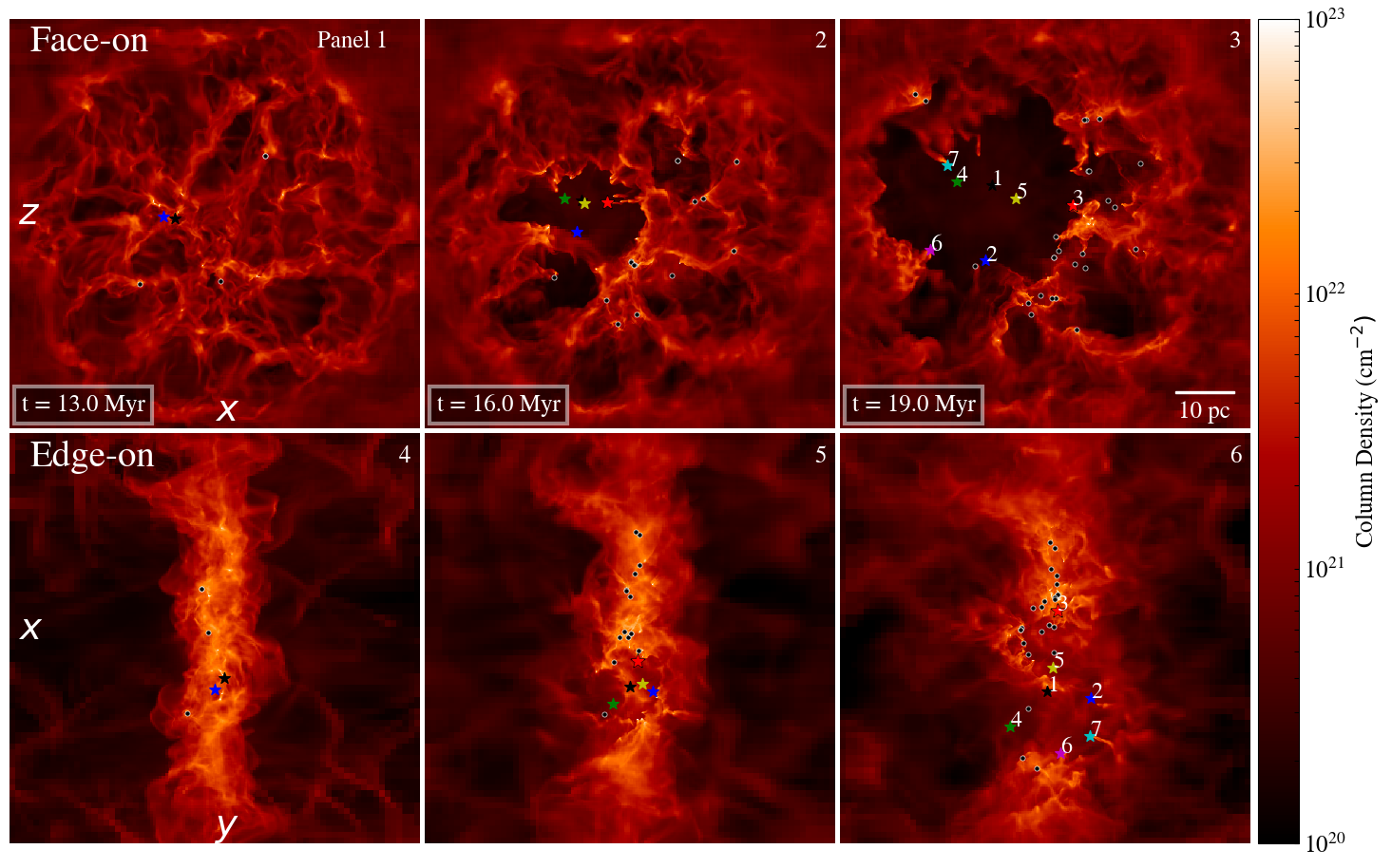} 
\caption{Column density maps for three different times (13, 16 and 19 Myr) showing face-on and edge-on views (upper and lower panels respectively). The coloured stars represent the projected position of the sink particles related with the H\,{\normalsize II} region we are analysing (other sinks are marked as black dots). The projections correspond to the 70 pc central sub-box.
 \label{fig:proj}}
\end{figure*}

The star formation process is modeled by dynamically refining in regions of high density and once we reach the maximum level of refinement allowed, a sink particle can be formed if the density in this cell exceeds a threshold number density, $n_{\rm thr} \simeq 4.2 \times 10^{6} \, \ppcc$, among other standard sink-formation tests \citep{Federrath+10}. Once the sink is formed, it can increase its mass via accretion from their surroundings. Given the size of our numerical box and the maximum resolution we can achieve ($\simeq 0.03$ pc), the sink particles rapidly reach hundreds of solar masses via accretion, and therefore we must not treat them as single stars but rather as a group or a small cluster of stars. Since the UV radiation flux will be dominated by the most massive star in the cluster, a standard, \citet{Kroupa01}-type initial mass function is assumed for each sink, in order to estimate the mass of the most massive star that the sink can host. Thus, the sink radiates according to this flux, which it is estimated from zero-age main sequence models at solar metallicity \citep{Paxton04}.

\begin{figure*}
   \centering
   \begin{minipage}{\textwidth}
     \centering
     \includegraphics[width=0.75\linewidth]{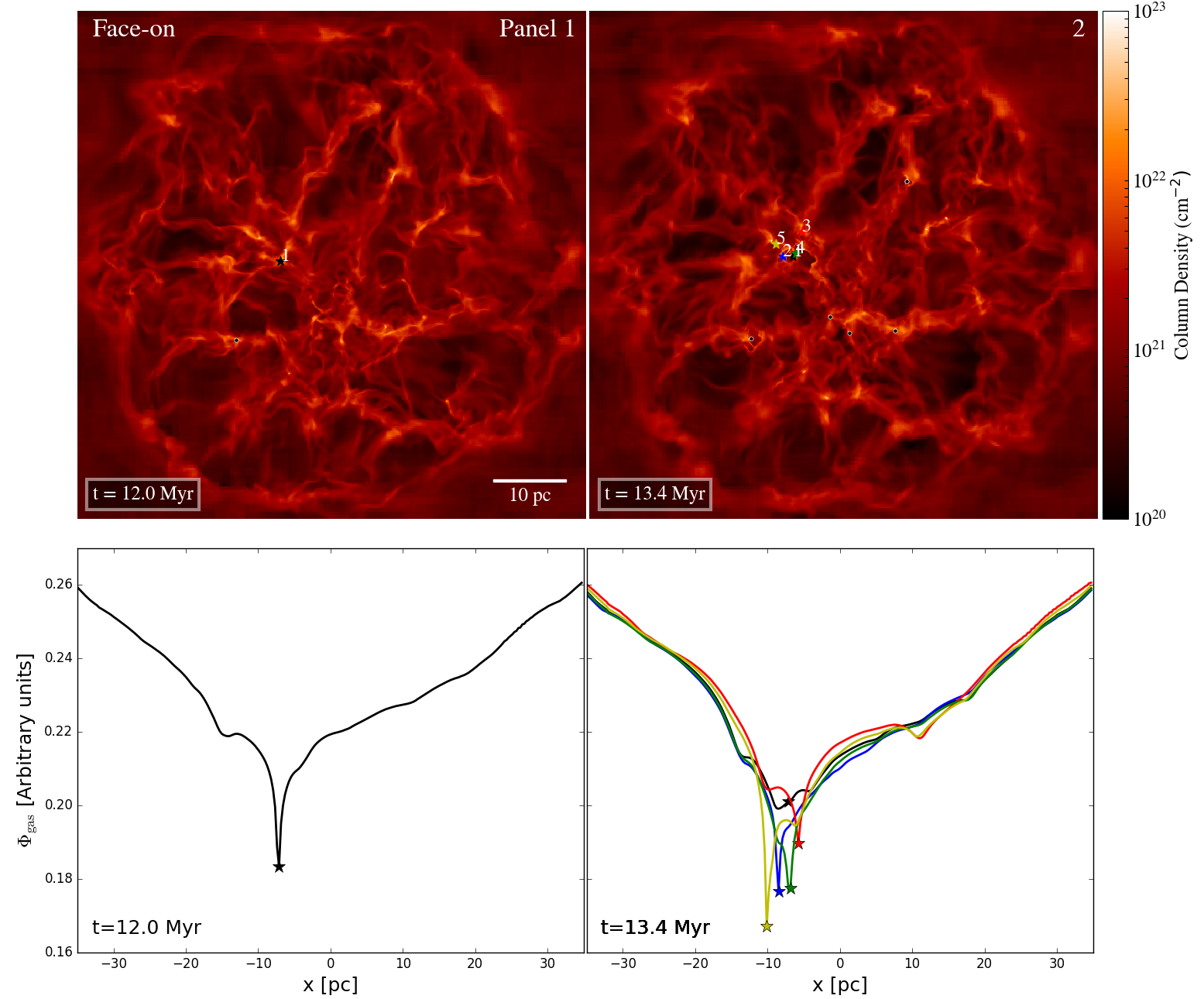}
    \end{minipage}\hfill
    \begin{minipage}{\textwidth}
     \centering
     \includegraphics[width=0.75\linewidth]{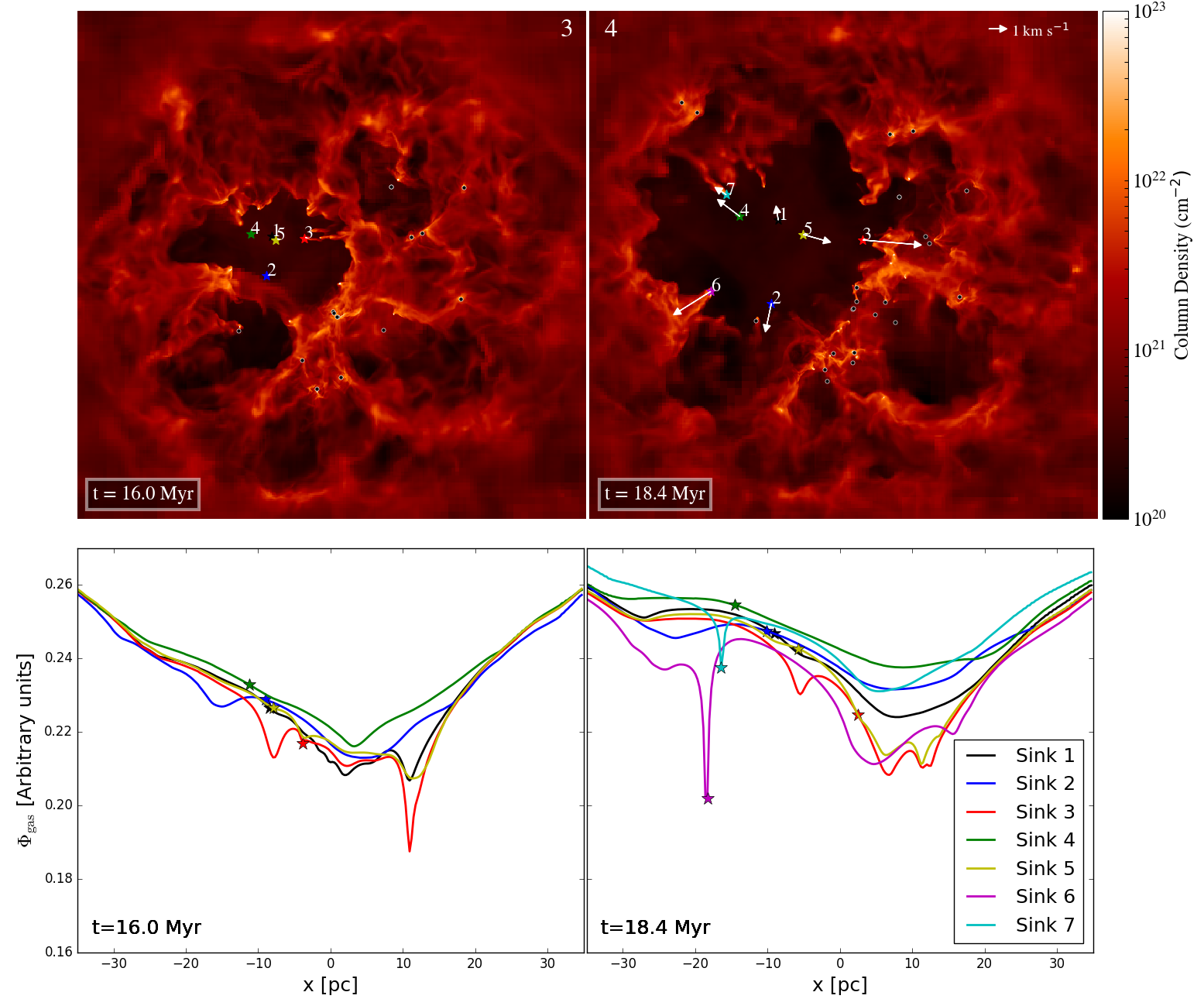}
    \end{minipage}
    \captionof{figure}{Face-on column density maps of the simulated cloud at $t=$12, 13.4, 16  and 18.4 Myr (panels 1-4). As in Fig. \ref{fig:proj}, the coloured stars represent the projected position of the sink particles. The arrows in panel 4 denote the projected velocity vectors (for reference, the 1~\kms ~vector is shown in the upper-right corner of the same panel). Panels below each column density map show profiles (along the $x$--direction) of the corresponding total gravitational potential (gas + sinks) in the frame of reference of each sink. See the evolution of this figure in a supplementary movie.}
    \label{fig:maps_evol}
\end{figure*}

Finally, the heating and cooling rates are estimated by breaking them into heating and cooling processes associated with ionisation of hydrogen atoms and the other thermal and chemical processes. For the former, the photoionisation heating rate is calculated by solving the radiative transfer equation \citep{Rijkhorst06,Peters+10}, whereas for the cooling
we consider ions-electrons collisions as the main mechanism for energy loss \citep[see, e.g.,][]{Dalgarno-McCray72}. On the other hand, for heating and cooling that are not directly due to ionisation, we use analytic fits by \cite{KI00,KI02}, which take into account heating by cosmic rays and cooling due to atomic and molecular lines and atomic and molecular collisions with dust, among others \citep[see also][]{Wolfire+95}. The MHD and the heating/cooling terms are coupled through operator splitting in the energy conservation equation as source terms. For further information about the numerical model we refer the reader to \citet{Rijkhorst06, Peters+10, ZA+18, ZA+19}.

\section{Results} \label{sec:results}

To illustrate the evolution of the simulated cloud, in Fig.~\ref{fig:proj} we show face-on (upper panels) and edge-on (lower panels) views of our simulated cloud, at three different times: 13 Myr (left panels), 16 Myr (middle panels) and 19 Myr (right panels).\footnote{These times refer to the total evolution of the simulation. However, it is worth noting that our cloud span a considerable fraction 
of its evolution in a cold-neutral phase \citep{VS+18} and it eventually become gravitationally unstable and starts forming stars at roughly the same time ($\sim$12 Myr). At this point the cloud may also be considered as molecular \citep[as proposed by ][]{HBB01}.}
The colour code denotes column density. Sinks in the expanding cluster, which will be analysed in more detail below, are drawn as coloured stars. Sinks that are outside the cluster are denoted with black dots. This figure shows the complex nature of the filamentary structure produced by the interplay of the non-homogeneous colliding flows that have suffered a phase transition and are already collapsing in a highly non-linear velocity field. In the evolutionary sequence (see the supplementary movie) it is interesting to notice that roughly 1 Myr after the first sink starts to radiate, the sink particles get an outward acceleration. This is specially true for sink 3 (red star), which at the beginning of the simulation is falling into the centre of the region, moving toward lower values of the $y$ (vertical) axis in the map, and changes its movement towards larger values of the $x$ (horizontal) axis.

In order to understand more deeply the nature of the motions of the sink particles,  Fig.~\ref{fig:maps_evol} shows a set of four double-plots panels, at times 12, 13.4, 16, and 18.4 Myr. In each panel, the upper map shows the face-on column map of our cloud, while the plot below shows cuts of the total gravitational potential (due to gas and sinks) along the horizontal axis, at the $(y, z)$ position of each sink in the map (in other words, each $x$ cut of the gravitational potential is made at the $(y, z)$ position of its corresponding sink.

In the upper left panel, we can see the morphology of the cloud when it has formed only the very first massive sink (with a mass of $\sim$~36~$M_\odot$), at $t=12$~Myr. At this moment, the cloud is falling into the centre of the box. From the corresponding gravitational field plot, it can be seen that the total gravitational potential has a minimum at the position of the sink, but also that it has some substructure, with a few other maxima and minima.

In the upper right map, at $t=13.4$~Myr, 5 sink particles have already formed with masses above a few times 10~$M_\odot$ each. It is important to remember that each sink continues accreting mass as the simulation evolves.  At this moment, they are relatively close to each other. Although it is not noticed in the map, the supplementary movie shows that, at this very moment, sink 3, for instance, is still falling in, while the expanding H\,{\normalsize II} region produced by sink 1 starts to blow-out the gas of the region. In the corresponding gravitational potential plot (upper-right panel), furthermore, it can be noticed that sinks 1 (black) and 4 (green) are not anymore in the minimum of the gravitational potential. Instead, the minimum is located elsewhere, and they are starting to change their direction of motion due to gravitational feedback produced by the mass that is continuously being redistributed. 

At later times, sink 3 is moving rightwards at larger velocities. This can be seen in the lower panels of Fig.~\ref{fig:maps_evol}. In fact, from the supplementary movie, it can be noticed that at earlier times, the sinks were originally approaching each other, but in the time lapse between 13.4 and 15.5~Myr, once enough mass is removed from the centre, they start flying apart from each-other. Since sinks only respond to the total gravitational field and conserve momentum from the surroundings when they accrete, the expansion of the cluster of sinks cannot be due to other effect than the gravitational field of the cloud, and thus, they should be getting a net acceleration from the outside.   

From the lower panels of Fig.~\ref{fig:maps_evol} we notice that sinks 1--5 are not in the deeper region of the gravitational field, since it has moved to the right. These 5 sinks are in a large-scale gradient of the gravitational field, feeling a net force to the right. Those sinks that were originally moving to the left (sinks 1,2,4), are already suffering a deceleration, while sinks that were moving originally to the right (sinks 3 and 5) continue increasing their velocity.

\begin{figure}
\includegraphics[width=1.0\hsize]{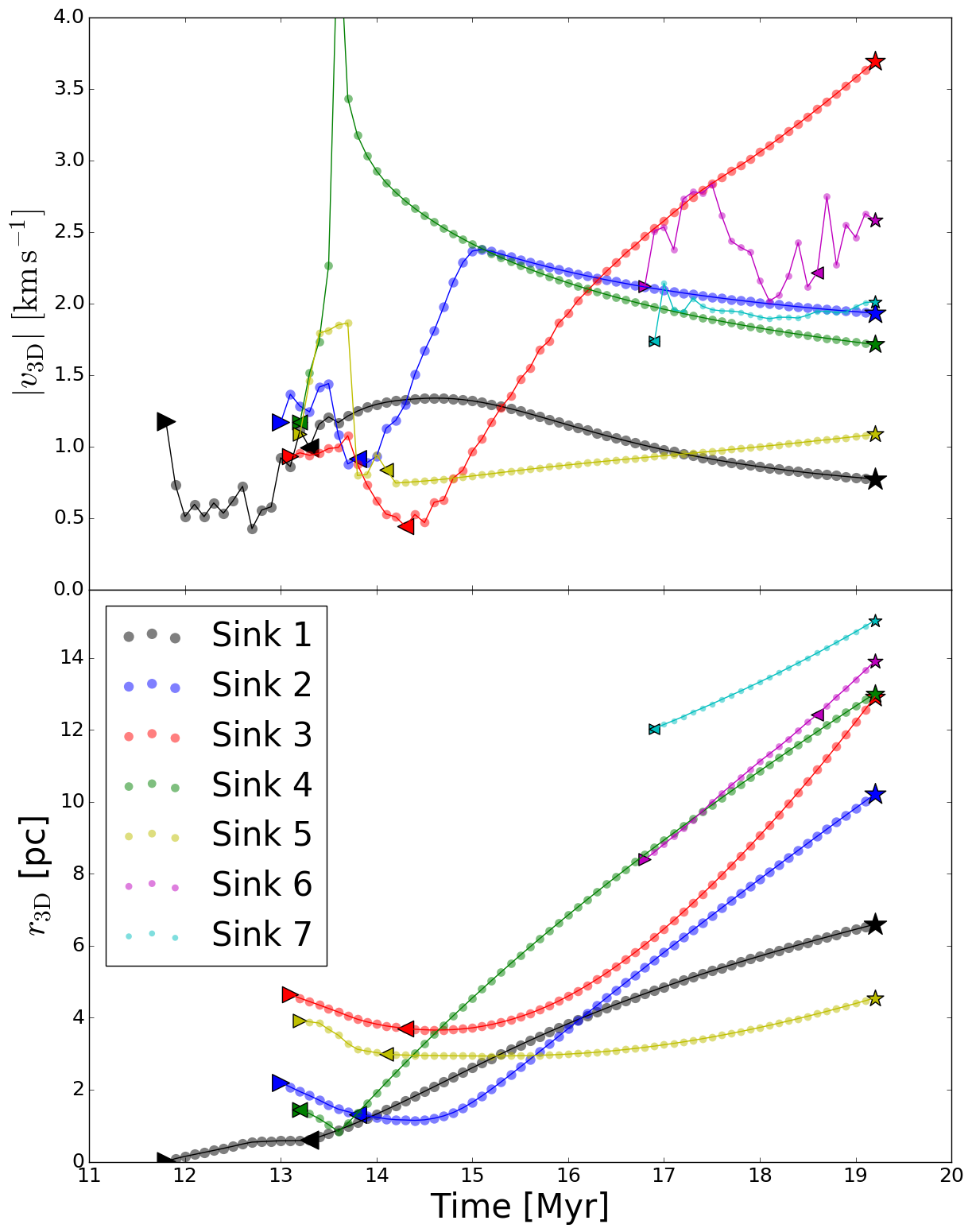}
\caption{Evolution of the velocity magnitude of each sink (top panel) and distance of each sink to the region centre (lower panel). For each sink, the starting time (zero age) is denoted by a right-pointing triangle symbol, the time at which the accretion stops by a left-pointing triangle symbol, and the final time of the simulation is indicated by a star symbol. Note that each sink is represented with a different colour according to Figs. \ref{fig:proj} and \ref{fig:maps_evol}. Each point corresponds to a single timestep in the simulation.
\label{fig:velocity-distance}}
\end{figure}

In Fig.~\ref{fig:velocity-distance} we show, as a function of time, the magnitude of the velocity vector, in the frame of the simulation (upper panel), and the distance from each sink to the position where the first sink was born (lower panel). We note three different stages in the evolution of the sinks (see upper panel of Fig.~\ref{fig:velocity-distance}). In the first stage, after a sink is created, its 3D velocity strongly oscillates. All our 7 sinks transit this phase. These changes in the velocities are a numerical effect, corresponding to the accretion phase: as has been shown in different numerical simulations, accretion flows are not necessarily symmetric, neither continuous \citep{Smith+11, BP+15, Kuznetsova+18}. They are intermittent, and occur preferentially through filaments. In the same figure, the left-pointing triangle symbols denote the moment when the accretion into the sink has stopped, i.e., when the gas in the vicinity of the sink has been blown out by the H\,{\normalsize II} region.

The second stage, experimented by sinks~1--5, consists of an acceleration process. In some cases, this process is fast (sinks~2, 4), moderate (sinks~1, 3), or slow (sink~5). The physical reason is the one we depicted before: the ``floor" of the gravitational potential changes, and the sinks transit from been in the local potential well, to be in a hillside of the potential cusp, for a substantial time in the evolution. This is the period in which the sinks acquire their expansion velocity. Finally, in the third stage, the acceleration changes. Three of the sinks (1, 2, 4) exhibit a period of deceleration, while sink~3 keeps accelerating, although at a lower rate. Sinks~6 and 7 never reach the 2nd or third stages, and sink~5 never reaches stage 3.

The reason for the sinks to be accelerated, regardless its rate of acceleration, is the same: the asymmetric mass distribution of the parental cloud makes a larger gravitational well on the right hand side of the H\,{\normalsize II} region compared to the left hand side. Thus, sinks moving to the left (1, 2 and 4) feel a force that opposes to their direction of motion, and thus decelerate, while sinks~3 and 5 feel a force in the same direction. Such acceleration and deceleration are seen in the supplementary movie related to Fig.~\ref{fig:maps_evol} and they are produced since the expansion centre of the  H\,{\normalsize II} region (and the associated cluster) is not at the centre of the molecular cloud and thus the global gravitational potential centre is elsewhere.  

The case of sinks~6 and 7 deserves special attention: they are located in a sort of ``pillars" produced by the expanding H\,{\normalsize II} region. These pillars are also being pushed away by the overpressure of the H\,{\normalsize II} region. However, sinks~6 and 7 never decouple from their pillar. This means that, as the pillar is pushed away, its local gravity pulls the sink it hosts, which results in a net outward acceleration of the stellar particles. 

Thus, while some of the sinks are pulled away by the overall gravitational field, other sinks are pulled by their local gravitational field, as they have not get rid of their parental clump. 

\begin{figure*}
\includegraphics[width=0.49\textwidth]{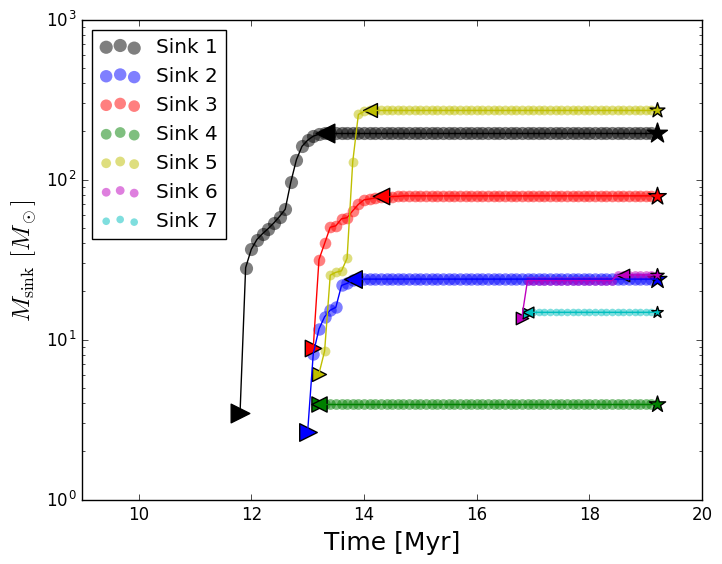} 
\includegraphics[width=0.49\textwidth]{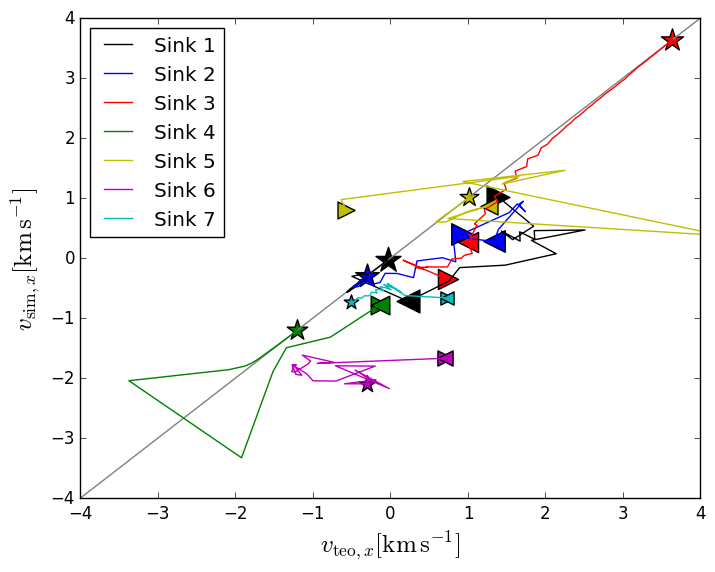}
\caption{{\it Left:} Mass evolution of each sink. {\it Right:} $x$ velocity ($v_{ {\rm sim}, x}$) measured directly from the simulation {\it vs.} the $x$ velocity calculated from the local gradient of the gravitational potential of the gas ($v_{{\rm teo},x}$). The symbols meaning is the same as in Fig. \ref{fig:velocity-distance}. See an animation of this figure in the supplementary material.
\label{fig:M_t_and_v3d_vnum}}
\end{figure*}

In the lower panel of Fig.~\ref{fig:velocity-distance}, the initially decreasing, and lately increasing distances between sinks~1--5 at their first stages of life, shows us that these sinks transit from a phase of contraction to a phase of expansion. Such change in their position is clearly seen in the supplementary movie associated to Fig. \ref{fig:maps_evol}, and emphasises the importance of the evolving gravitational potential, which moves its well from the innermost part of the cluster, to the external part.

In the left panel of Fig.~\ref{fig:M_t_and_v3d_vnum} (see also the supplementary animation of this plot) we show the mass history of the sinks. The symbols in these plots are the same as those in  Fig.~\ref{fig:velocity-distance}. As can be seen, the sinks rapidly increase their mass, and then, they stop accreting. This is because the H\,{\normalsize II} region expands and swipes away all the dense gas. Additionally, we notice that the initial velocity oscillations discussed in Fig.~\ref{fig:velocity-distance} correspond to the epoch of accretion.  
In the right panel we plot the actual $x$ velocity ($v_{ {\rm sim}, x}$), against the velocity theoretically expected from the gradient of the gravitational potential of the gas ($\nabla\phi_{\rm gas}$), acting as if the potential was constant during the time interval between dumps of the simulation ($\Delta t$), i.e., $v_{\rm teo} \equiv v_0 + \nabla\phi_{\rm gas} \ \Delta t$. 
In this case, if the velocity of the sinks were only defined by the gravitational field of the gas, and if the field evolved slowly enough to be considered constant during $\Delta t$, sinks would have to move along the identity line. As can be seen, during the first stages of the existence of the sinks, the numerical and predicted velocities are different. As mentioned above, during this period of time, asymmetric accretion modifies the momentum of the sinks strongly. However, it is worth noticing that as soon as the accretion stops, all but sinks~6 and 7 move (and they have to) along the identity line, reinforcing the idea commented before that the strong changes in velocity seen in Fig.~\ref{fig:velocity-distance} are due to momentum imprint from asymmetrical accretion, which in turn is due to the limited resolution of the simulation. 

Again, the cases of sinks~6 and 7 are worth mentioning independently. These sinks are created at a later time in the simulation, at the cusps of pillars created by the expansion of the H\,{\normalsize II} region, and they move along with the pillar. In these two cases, the predicted-vs-numerical velocity curve (right panel in Fig.\ref{fig:M_t_and_v3d_vnum}) does not follow the identity line. Instead, they jump from one value to another. The initial phase can be attributed to the asymmetric accretion/resolution issue mentioned above. However, this behaviour continues even after the accretion stops. This second effect is numerical: the variation of the gravitational potential in these places is quite fast, and we calculated the predicted velocity ($x-$axis in Fig.~\ref{fig:M_t_and_v3d_vnum}) using data-dumps that were spaced 0.1~Myr in time, much larger than the typical variation in time of the gravitational potential of the pillar. As a consequence, by taking the acceleration at a given time, $\nabla \phi_{\rm gas}(t)$ to be the typical acceleration of the whole period of time $\Delta t$, we have made a crude estimate of the actual mean acceleration during the same period of time. This produces the observed lack of correspondence between the numerical and the predicted velocities of those sinks.

Even though the resulting numerical value of the velocity of sinks~6 and 7 is not precise, what is important in this case is that the sink moves with the velocity of the gas, and the gas itself is gravitationally pulling the sink with it. This suggests that clusters like {B30} or B35, which are located in the cusps of pillars in $\lambda$-Ori, might be moving along with their own pillar, and might have been pulled by the gravity of the pillar itself. If the evolution of clusters embedded in gas pillars is directly linked to the evolution of the pillars themselves, then this puts some additional interest to some systems. For instance, the DR15 cluster in Cygnus-X is suspected to have a relatively slow gas dispersal respect to the evolution of its young stellar members in the absence of a massive star \citep{rivera15}. This case contrasts with massive member hosting clusters like NGC 3603, whose pillars appear to suffer strong photoevaporation \citep{wmoq13}; it will be important to know what is the contribution of the internal kinematics of the cluster-pillar system that adds to the feedback contribution from massive members and erosion from external radiation. We plan to study this in a further contribution.

\section{Discussion: A comprehensive scenario for the formation and evolution of young clusters.}\label{sec:discussion}

Although the intention of the present contribution is not to reproduce in detail the kinematical and morphological features observed in $\lambda$-Ori, this region draw our attention to analyse the possible relevance of the gravitational potential once it is flipped-up as a consequence of the inside-out removal  of the dense gas by the H\,{\normalsize II} region. 

Giant molecular clouds (GMCs), the sites where stars are born, are formed from large-scale flows \citep[in the form of spiral arms, expansion waves, or large-scale gravitational instability producing converging flows of the atomic gas clouds, see e.g., ][and references therein]{Dobbs+14}. In either case, as the diffuse gas is compressed, it cools down, undergoing thermal instability. As a result, the originally diffuse, warm atomic gas at temperatures of several \diezala 3 K cools down to a few times 10~K \citep{HennebellePerault99, Heitsch+06,VS+07}. As the gas becomes more dense and opaque to the dissociating UV radiation, it also turns into the molecular phase. As a result, the compressed region, which we now call molecular cloud, becomes rapidly gravitationally unstable and collapses to form stars within a few million years \citep{HBB01}. All this process may accumulate enough mass to form not one or a few, but hundreds or thousands of stars in compact stellar clusters. 

Numerical simulations have shown that the first generation of massive stars formed in a GMC could in principle disperse out a significant fraction of their parental gas, limiting the amount of gas available for star formation to about a few percent of the  mass of the cloud \citep{Dobbs+14}. Different studies have focused on the identity and survival of  newborn stellar clusters as the gravitational potential of the gas that formed them is lost (due to the feedback of the massive stars). The actual fate of the newborn stellar clusters strongly depends on when, and how rapidly, the gas in the parental cloud is lost \citep{Smith+13,Farias+18}. The prescription for the gas removal has been modeled numerically with relative success, first, by imposing an artificial gravitational potential, instead of one that is consistent with the distribution of mass of the parental cloud. Second, and probably more important, by either assuming that the potential vanishes instantaneously \citep{Farias+18}, or preserves its original shape, but it gently
lowers its depth in an arbitrary timescale \citep{Smith+13}. As a result, none of these prescriptions account for the fact that the mass of the expelled gas (which is actually  larger by a factor of at $\geq 3$, given the efficiencies) removes in a continuous way the central potential well while it is creating or enhancing new local wells outside, such that the newly born clusters will be gravitationally pulled away, towards the edges of the bubble. In other words, most of the mass of the cloud is not disappearing in an arbitrary way, but it is instead being moved at a finite velocity, from the central collapsing regions to the periphery. 

The present contribution shows that the transport of mass from the inside of the core to the outside of the H~II region may have a significant gravitational influence on the dynamics and kinematics of the cluster. The bottom of the potential well at the centre of the cluster transitions from having a valley at the time of the cluster formation, to having a cusp or peak by the time the shell is formed around the H\,{\normalsize II} cavity. As the fraction of mass that it is converted into stars in a GMC is small \citep{LadaLada03}, it can then be expected that the gravitational pull from the outside might be important and the kinematics of the clusters could be substantially affected in non-symmetric configurations. This mechanism can be expected given the fractal nature of clouds \citep[e.g., ][]{Scalo90, Falgarone+91}.

Our simulations show that, indeed, this is the case, and that some sort of gravitational feedback occurs in expanding H\,{\normalsize II} regions: the dense gas is expelled by the H\,{\normalsize II} region, and the stars are then pulled out by the gravitational field of the gas.
This effect occurs mainly in two different ways: either the stellar cluster is seated on the hillside of the flipped-up gravitational field (sinks~1--5), or it is seated in the local potential well of a dense pillar which is anyway pulled away (sinks 6, 7).

Although we have only few (7) sinks which, given their masses, they should be interpreted as small stellar clusters (5 within the H\,{\normalsize II} region and 2 on the pillars). The nature of the gravitational feedback imprinted by the gas on these few groups of stars is asymmetric, since each sink is pulled-out in a different direction. Interestingly, Gaia DR2 results reported by \citet{Wright+19} found an asymmetrical expansion of the NGC~6530 cluster in the Lagoon Nebula. They discuss how non-symmetrical residual gas expulsion via feedback is normally modelled as symmetrical, which is not in agreement with their observations. Our simulations, thus, are consistent with an scenario where non-symmetrical gas expulsion can lead to an also non-symmetrical expansion of the cluster, and thus, are also consistent with the observations reported by \citet{Wright+19}.

As we mentioned before, our simulated clouds consist of a network of highly structured and not homogeneous filaments that naturally forms by converging flows and that eventually enters in a state of {\it global hierarchical collapse}. Thus, the cloud inherit part of their kinematics to the sinks (sub-clusters) and once the first massive sinks begin to radiate, the dense gas around the cluster expands also in an non homogeneous way imprinting thus a net force to the sinks.

\subsection{{Limitations}} \label{subsec:limitations}

As commented in \S\ref{sec:methods}, the initial conditions are two converging streams of diffuse gas. As a consequence, our cloud is flattened, and thus, a natural source of concern could be whether the particular geometry of our simulation plays a role enhancing the relevance of the external gravitational field. 

A detailed analysis of different configurations and the frequency of this effect is scope of an forthcoming contribution. Here, we want to stress the possibility that the gravitational feedback pulling out the stars can play a role, and that it is a mechanism that needs to be reviewed in detail.

In any event, regarding geometry, we believe it is important to note that indeed, the initial conditions of our simulation is highly idealised. However, the simulated cloud is not that unrealistic: the compressed sheet is a $\sim $10~pc wide, 50~pc long filament when it is seen edge-on, typical values of the dimensions of MCs. It is not clear  whether the observed filamentary nature of MCs is a consequence of clouds been a network of inhomogeneous sheets, some of them intersecting to each other, some others seen edge-on, and whether those are actual filaments. For instance, the Musca cloud has a sheet-like morphology viewed edge-on \citep{Tritsis+18}. Similarly, recent kinematic study of the B213 filament in Taurus suggests that it can also be the dense part of a sheet produced by a large-scale compression \citep[][]{Shimajiri+19}.
  
Besides the fact that clouds are far from being spherical, stellar clusters hardly appear at their centre. On the contrary, they frequently appear at the tips or edges, or forming roughly at the same time on the densest parts of irregular clouds \citep[as in the Rosette MC; e.g.,][]{Ybarra+13}, probably as a consequence of gravitational focusing of regions with low curvature radius \citep{HartmannBurkert07, Kuznetsova17}. Thus, our idealised configuration may play a role, as well as any other configuration, and thus, this has to be addressed by means of analysing different simulations.

In addition, the numerical simulations presented have some limitations that are worth mentioning:

\begin{itemize} 

\item{} Due to the limited resolution of the simulation at small scales, accretion events tend to be more massive and intermittent than what they could be in nature, and we have shown that this imprints strong changes of momentum over the sink. Thus, the numerical velocity cannot be taken at face value when the sinks are still accreting gas from their surroundings. 

\item{} Also, in actual star forming regions, although asymmetric accretion flows might be present, they not necessarily reach  the star; instead, they might create density enhancements around its vicinity. Such features have just started to be observed recently with high resolution observations from the Atacama Large Millimeter/submillimeter Array \citep[ALMA;][]{Tokuda+18}. 

\item On the other hand, due to numerical limitations, we cannot resolve the fate of individual stars that composes the clusters, and then, detailed kinematics of each group cannot be followed. Further numerical simulations should be performed in this regard. 
\item{} {Finally, although we demonstrate that the ionising feedback is able to produce a 20~pc bubble, still it is necessary to quantify the effects of other kinds of feedback, e.g., radiation pressure, winds, supernovae explosions, etc. Clearly, all of these mechanisms can affect the evolution of the dynamics of gas, and it will be necessary to perform detailed calculations in order to really understand the nature and relevance of the ``gravitational feedback".}

\end{itemize}

\section{Summary and concluding remarks}\label{sec:summary}

We present a simple but novel mechanism that could imprint substantial dynamical effects on the stars in young stellar clusters. Our work was  motivated by the morphology and kinematics of the $\lambda$-Ori star forming complex, and uses numerical simulations to study the complete evolution of massive star-forming molecular clouds. The clouds form by compressions in the warm neutral medium and via accretion they eventually acquire enough mass to enter in a state of {\it global hierarchical collapse}. When massive stars appear, they start ionising and dispersing the parental cloud. We study the dynamical effects of the gravitational potential of the expanding gas on the sink particles, which represent stellar clusters. 

Our results show  that in order to understand the present dynamics of young stellar clusters, it is important to consider the evolution of a realistic gravitational potential in which the mass of the parent cloud is considered. Simple prescriptions for the molecular cloud gravitational potential diminishing its relevance in time will not capture the essence of this mechanism, since the redistribution of the mass in an inherently asymmetric way from the inner parts of the core to the outside is not taken into account.

We find that clouds in {\it global hierarchical collapse} can reproduce the dynamics of observed expanding young stellar clusters. In this framework, our model shows that while the stars might be falling into the gravitational potential in the early stages of formation, once massive stars start emitting ionising radiation, an expanding bubble is generated, moving the internal mass outside the bubble, and flipping-up the gravitational potential well. This imprints an outward acceleration to the sinks, which experiment a ``Hubble-like" expansion due to the attraction of this well. 

We have shown for first time that the dispersal of Galactic stellar clusters can be triggered by the change in the gravitational potential well, which is dominated by the mass of the cloud. This mechanism has not been considered before, mainly because in previous, more simplified models, the gravitational potential was assumed only to disappear, and thus, it was assumed not to do an additional work in pulling out the stars. 

\section*{Acknowledgments}
We thank the referee for helpful comments and suggestions. M.Z.A. acknowledges support from CONACYT grant number A1-S-54450 to Abraham Luna Castellanos (INAOE). J.B.P., J.H., and C.R.Z. acknowledge support from program UNAM-DGAPA-PAPIIT grants IN111219, IA103017, and IA102319, respectively. This work has made use of data from the European Space Agency (ESA) mission {\it Gaia (\url{https://www.cosmos.esa.int/gaia}}), processed by the {\it Gaia} Data Processing and Analysis Consortium (DPAC, \url{https://www.cosmos.esa.int/web/gaia/dpac/consortium}). Funding for the DPAC has been provided by national institutions, in particular the institutions participating in the {\it Gaia} Multilateral Agreement. The visualisation and analysis were carried out with the {\tt yt} software \citep{yt}. The authors thankfully acknowledge computer resources, technical advise and support provided by {\it Laboratorio Nacional de Superc\'omputo del Sureste de M\'exico} (LNS), a member of the CONACYT network of national laboratories.

\bibliographystyle{mnras}
\bibliography{references}

%
%
%
\bsp	
\label{lastpage}
\end{document}